# Why the 30% Mansfield Rule can't work:
## *A supply-demand empirical analysis of leadership in the legal profession*
*Paola Cecchi Dimeglio*[1]
*Harvard Law School*
*Harvard Kennedy School*

## I. Introduction

Across the legal profession, statistics related to the numbers of women and other underrepresented groups in leadership roles continues to paint a bleak picture of diversity and inclusion.[2] Women hold 19% of these roles, and racial minorities hold 6.6%. LGBTQ± individuals occupy fewer than 3% while differently abled people, as defined by the American Disabilities Act, account for under 1%.[3] Industry-wide, this absence of diversity has become the defining rule.[4] Numbers remain low and stagnant, despite efforts and campaigns aimed at improving inclusion of women and underrepresented groups at leadership levels in the legal profession.[5]

Some approaches to closing this gap have focused on the cause; some have devised and applied solutions.[6] These drivers and cures are distinct, and one does not necessarily contain meaningful information about the other.[7] Questions about the efficacy of many of these solutions remain essentially unanswered.[8] The bulk of what has been devised and applied has not been evidence-based or subject to scientific rigor. It has also not accounted for the manner in which the supply of lawyers impacts the feasibility or effectiveness of diversity efforts, particularly at the level of leadership.[9]

---

[1] Paola Cecchi-Dimeglio, Ph.D., J.D., LL.M., Ms.C.. She can be reached at pcecchidimeglio@law.harvard.edu. Acknowledgement and appreciation are extended to many members of the Harvard Law School, Harvard Kennedy School, and Harvard Business School and to colleagues at other universities especially for their faculty scholarship and feedback. The author is in particular thankful to Bruce Green, David Wilkins, Scott Cummings, David Luban, Carole Silver, Swethaa Ballakrishnen, Renee Knake, Michele DeStefano, Russ Pearce, Leslie Levin, Atinuke Adediran and Peter Kamminga for their helpful comments, guidance, assistance, and support.

[2] Deborah L Rhode, *Diversity and Gender Equity in Legal Practice*, 82 UNIVERSITY OF CINCINNATI LAW REVIEW 31 (2018); Deborah L. Rhode, *From platitudes to priorities: diversity and gender equity in law firms*, 24 THE GEORGETOWN JOURNAL OF LEGAL ETHICS 1041 (2011); HANDBOOK OF THE PSYCHOLOGY OF WOMEN AND GENDER, (Rhoda K. Unger ed., 2004).

[3] MCCA, *2020 VAULT/MCCA LAW FIRM DIVERSITY SURVEY REPORT*, (2020), https://www.mcca.com/wp-content/uploads/2021/05/2020-Vault_MCCA-Law-Firm-Diversity-Survey-Report-FINAL-R2.pdf; NAWL, *2020 Survey report on the promotion and retention of women in law firms*, 42 (2020), https://www.nawl.org/p/cm/ld/fid=2019.

[4] Atinuke O. Adediran, *The journey: moving racial diversification forward from mere commitment to shared value in elite law firms*, 25 INTERNATIONAL JOURNAL OF THE LEGAL PROFESSION 67 (2018); DIVERSITY IN PRACTICE: RACE, GENDER, AND CLASS IN LEGAL AND PROFESSIONAL CAREERS, (Spencer Headworth et al. eds., 2016); David B. Wilkins, *Identities and Roles: Race, Recognition, and Professional Responsibility*, MD. L. REV. (1998), https://dash.harvard.edu/handle/1/8789617 (last visited Jun 12, 2021); D. B. Wilkins & G. M. Gulati, *Why Are There So Few Black Lawyers in Corporate Law Firms? An Institutional Analysis*, 84 CALIFORNIA LAW REVIEW 493 (1996).

[5] ROBERT L. NELSON, PARTNERS WITH POWER 202 (2020), http://www.degruyter.com/document/doi/10.1525/9780520317970/html (last visited Sep 30, 2022); Rhode, *supra* note 2; Marjorie Rhodes, Sarah-Jane Leslie & Christina M. Tworek, *Cultural transmission of social essentialism*, 109 PROCEEDINGS OF THE NATIONAL ACADEMY OF SCIENCES - PNAS 13526 (2012); Elizabeth H. Gorman & Fiona M. Kay, *Racial and ethnic minority representation in large U.S. law firms*, in STUDIES IN LAW, POLITICS, AND SOCIETY 211 (Austin Sarat ed., 2010), https://www.emerald.com/insight/content/doi/10.1108/S1059-4337(2010)0000052010/full/html (last visited Jun 12, 2021); Crystal Hoyt, *Women, Men, and Leadership: Exploring the Gender Gap at the Top*, 4 SOCIAL AND PERSONALITY PSYCHOLOGY COMPASS 484 (2010); Carrie Menkel-Meadow, *Exploring a Research Agenda of the Feminization of the Legal Profession: Theories of Gender and Social Change*, 14 LAW & SOCIAL INQUIRY 289 (1989).

[6] Paola Cecchi-Dimeglio, *Designing Equality in the Legal Profession: A Nudging Approach*, 24 HARVARD NEGOTIATION LAW REVIEW 24 (2019).

In addressing the underlying mathematical reality that can cause these DEI efforts to falter, I undertake two efforts:
1. Laying out the current legal profession supply and demand landscape at the leadership level, as this is essential to any empirical consideration of what works or fails with respect to DEI in the profession.
2. Addressing some of the thinking that has been utilized to create the illusion of progress and explain why it cannot produce the claimed results.

Scholarship and research in the area of labor and economics highlight the significance of supply-demand dynamics in labor markets.[10] Understanding these dynamics enables practitioners to predict and prepare for difficulty or disruption. It enables them to craft and apply policies that foster change, target demographics, and influence supply. The supply-demand approach also enables accurate assessment of whether interventions succeeded or not.

These forecasting models and practices have been applied across many industries but have not been recently used in the legal profession.[11] The labor component of the legal industry consists of institutional, educational, financial, and other elements.[12] These parts interactively shape and influence supply and demand in the profession. Given the different set of moving parts that shape the labor force in the legal sector, effectively shifting demographics becomes a challenging effort, subject to multiple variables.

The empirical study presented here represents one of the first of its kind. Studies of the legal profession have not focused on the dynamics of supply and demand in the context of leadership positions (counsel and partner (equity and non-equity). Neither have they examined the interrelationships of these dynamics to race and gender demographic factors (white female and minorities (male and female).

In order to roundly approach and address these factors, this research examined the claims of success promoted by the Diversity Lab.[13] Specifically, it investigated the validity of claims that adopting a rule requiring that candidate pools consist of 30% women, racial minorities, or other underrepresented individuals was sufficient to increase the numbers of these groups hired or promoted into leadership roles at law firms. [14]

---

[7] Paola Cecchi Dimeglio & Hugh Simons, *LATERAL MOVES: AN EMPIRICAL INVESTIGATION OF CYCLICALITY, DIRECTIONAL MOBILITY, AND 5-YEAR RETENTION RATE BY GENDER AND AGE COHORT*, 42 J. LEGAL PROF. 27 (2018).

[8] Paola Cecchi Dimeglio, *Is the Mansfield Rule moving the needle for women and minorities?*, JOURNAL OF PROFESSIONS AND ORGANIZATION joac007 (2022).

[9] DEBORAH L. RHODE, LEGAL ETHICS (Seventh edition. ed. 2016).

[10] N. Fuchs-Schündeln & T.A. Hassan, *Natural Experiments in Macroeconomics*, 2 *in* HANDBOOK OF MACROECONOMICS 923 (2016), https://linkinghub.elsevier.com/retrieve/pii/S1574004816000094 (last visited Sep 30, 2022).

[11] H. C. Horack, *Supply and demand in legal profession*, 14 AMERICAN BAR ASSOCIATION JOURNAL 567 (1928).

[12] Swethaa Ballakrishnen, Priya Fielding-Singh & Devon Magliozzi, *Intentional Invisibility: Professional Women and the Navigation of Workplace Constraints*, 62 SOCIOLOGICAL PERSPECTIVES 23 (2019); Russell G. Pearce, Eli Wald & Swethaa S. Ballakrishnen, *Difference Blindness vs. Bias Awareness: Why Law Firms with the Best of Intentions Have Failed to Create Diverse Partnerships Colloquium: The Challenge of Equity in the Legal Profession: An International and Comparative Perspective*, 83 FORDHAM L. REV. 2407 (2014); Ethan Michelson, *Women in the Legal Profession, 1970-2010: A Study of the Global Supply of Lawyers*, 20 INDIANA JOURNAL OF GLOBAL LEGAL STUDIES 1071 (2013).

[13] John Lino, Jim Sandman & Caren Ulrich Stacy, *Diversifying Leadership: How the Mansfield Rule Is Driving Change*, (2022), https://news.bloomberglaw.com/us-law-week/diversifying-leadership-how-the-mansfield-rule-is-helping (last visited Sep 26, 2022).

The research presented here seeks to determine the supply-demand position of leadership in the legal profession and to establish whether market equilibrium exists for these counsel and partner roles. In its approach, the model considers the overall population then examines the two subdivisions of White female and minorities (female and male).

In order to frame supply and demand, the model determines the available population of lawyers and establishes the demand based on positions and job level. This action of comparing available candidates to positions permits determination of equilibrium in supply and demand. Are there enough candidates for a particular position or level? What is the percentage of surplus or shortage above or below equilibrium? As will be seen, insufficient supply can render percentage-based DEI efforts inoperable.

This article answers those questions for the period from 2017 to 2021, examining the supply-demand dynamics for the identified populations. With respect to the specific talents, the article first examines the broad population then drills down into underrepresented populations (White/Caucasian females, and minorities, female and male). It relies on data primarily from AML and utilizes other data from publicly available sources. These sources include the US Bureau of Labor Statistics, US Census Bureau, and the American Bar Association.

327 law firms comprise the data set of this study with a total population of over 248,628 lawyers. Female lawyers represent 37% of this total or 90,891 lawyers. Males represent 63% or 157,737 lawyers. The White/Caucasian component is 82% or 204,802 lawyers, and the minority component is 18% or 43,826 lawyers. While percentages are comparable to other data sets, the sample set used for this article is twice as large.

The analytical model examines if the existing supply and demand of lawyers for leadership positions, including counsel and partner (non-equity and equity partner), meet the numbers required to provide the 30% essential to the Mansfield rule, and determines whether there is equilibrium, surplus, or a shortage, which would make it impossible to operate the rule. The factors applied include supply and demand of lawyers at 30% for the underrepresented population versus the number required to provide 30% of that population in the candidate pool. Equilibrium, surplus, or shortage is thus determined.

This article is organized into five main parts: Part I provides an overview of the relevant literature; Part II presents the methodology; Part III provides an overview of the data; Part IV presents the results, and Part V delivers a discussion and concluding commentary.

## II. Literature review

### A. Supply & demand

Legal labor markets, like markets for other goods and services, demonstrate a supply and demand curve.[15] Common concerns, such as pay across education and demographic groups and levels of employment, impact supply-demand dynamics. Other impacting factors include government policies and the behaviors of firms or labor unions[16].

---

[14] Reference is made to numbers published at the Diversity Lab website and mentioned in a non-peer review article authored by the Diversity Lab; specifically, the author has not been able to reproduce the claimed results while meeting an acceptable and reliable scientific threshold required to validate a scientific claim.

[15] Dale W. Jorgenson et al., *U.S. Labor supply and demand in the long run*, 30 JOURNAL OF POLICY MODELING 603 (2008).

In the labor market, supply-side factors include the available population of workers, their level of education, the skills they offer, and their demographic composition.[17] The demand side consists of the jobs defined and offered by employers. Specifically, demand consists of positions at law firms and in-house roles at public and private organizations, the skills sought, compensation offered, and aspects of recruitment.[18]

Several factors and market forces impact the demand side, including globalization and automation. Strategic shifts, such as the use of contractor status lawyers over fulltime employees, and government regulations such as minimum wage also shape demand-side dynamics.[19]

Over the last 30-40 years, the interplay of market and institutional factors have produced observable shifts in the legal labor market.[20] Although they agree on this point, legal labor economists do not always agree on how much or how exactly these dynamics have shaped outcomes. There is more silence and less agreement regarding the impact these forces will have on employment and pay going forward, especially with respect to demographics (gender and race).

### B. The rise of the Mansfield Rule

The legal profession's Mansfield Rule modified and mimicked the NFL's Rooney Rule. Established in 2003, the latter sought to affect inclusion by requiring that at least one minority candidate be interviewed for any vacant head coach position. Despite a 2015 Cynthia DuBois study asserting the Rooney Rule's success, in 2020, there were less than five African Americans among the NFL's 32 head coaches. This gap has resulted in the assessment that Rooney is an ineffective "checkbox exercise".[21] What is often missed is that when such rules are applied and there is one diverse candidate in the pool, choice may be forced or appear to be forced on the basis of demographics, and a backlash of resentment may ensue. "Such policies may backfire…, even if minorities are preferred to non-minorities and are, on average, at least as qualified".[22]

Emerging from the 2016 Women in Law Hackathon hosted by Diversity Lab, in collaboration with Bloomberg Law and Stanford Law School (Diversity Lab 2016), the Mansfield Rule sought to move the needle on women and minorities in leadership positions at law firms.[23] Named for the first woman admitted to the practice of law in the United States, Arabella Mansfield, the rule asserts that the inclusion of 30% underrepresented individuals in the candidate pool will produce the DEI outcomes sought.

---

[16] Tsunao Okumura, *Nonparametric estimation of labor supply and demand factors*, 29 JOURNAL OF BUSINESS & ECONOMIC STATISTICS 174 (2011).

[17] H. BUNZEL, STRUCTURAL MODELS OF WAGE AND EMPLOYMENT DYNAMICS (1st ed. ed. 2006).

[18] JOHN QUIGGIN, ECONOMICS IN TWO LESSONS: WHY MARKETS WORK SO WELL, AND WHY THEY CAN FAIL SO BADLY (2019).

[19] MARIA REGINA REDINHA, MARIA RAQUEL GUIMARÃES & FRANCISCO LIBERAL FERNANDES, THE SHARING ECONOMY: LEGAL PROBLEMS OF A PERMUTATIONS AND COMBINATIONS SOCIETY (2019).

[20] Daron Acemoglu & David Autor, *Chapter 12 - Skills, Tasks and Technologies: Implications for Employment and Earnings*, in HANDBOOK OF LABOR ECONOMICS 1043 (David Card & Orley Ashenfelter eds., 2011), https://www.sciencedirect.com/science/article/pii/S0169721811024105 (last visited Sep 30, 2022).

[21] Norman Chad, *Perspective | The dearth of black coaches in the NFL is a problem that somehow still hasn't been fixed*, WASHINGTON POST, January 5, 2020, https://www.washingtonpost.com/sports/the-dearth-of-black-coaches-in-the-nfl-is-a-problem-that-somehow-still-hasnt-been-fixed/2020/01/05/5993904e-2e79-11ea-bcb3-ac6482c4a92f_story.html (last visited Jun 14, 2021); Cynthia DuBois, *The Impact of "Soft" Affirmative Action Policies on Minority Hiring in Executive Leadership: The Case of the NFL's Rooney Rule*, 18 AMERICAN LAW AND ECONOMICS REVIEW 208 (2016).

[22] Daniel Fershtman & Alessandro Pavan, *"Soft" Affirmative Action and Minority Recruitment*, 3 AMERICAN ECONOMIC REVIEW: INSIGHTS 1 (2021).

[23] Diversity Lab, *Hackathon Series « Diversity Lab*, (2016), https://www.diversitylab.com/diversity-in-law-hackathons/ (last visited Jun 14, 2021).

Following the implementation of the Mansfield Rule in 2017, participating law firms have sought to include 30% women and minority attorneys, and more recently LGBTQ± lawyers, in their candidate pools for leadership and governance roles, equity partner promotions, and lateral positions.[24] Per the rule, if a firm has identified a list of five candidates for an applicable opening, two of these candidates must be diverse.

Mansfield applies to open leadership and governance positions and internal promotion processes. Its scope includes "Equity partner promotions, Lateral partner and Senior Associate hiring searches and openings, election or appointment to practice group and office head leadership positions, election or appointment to Management/Executive Committee and/or Board of Directors, election or appointment to Partner Promotions/Nominations Committee, election or appointment to Compensation Committee, election or appointment to Chairperson and/or Managing Partner, participation in formal client pitches and transparent job responsibilities and processes for governance appointments/elections."

Overall, the Mansfield Rule aims to increase representation of women and minority groups in leadership roles at law firms by diversifying the candidate pool. Firms that adhere to the rule become "Mansfield Certified" the following year. There have been five generations of Mansfield certification: Pilot (07/2017-07/2018), Certified 2.0 (07/2018-07/2019), Certified 3.0 (07/2019-07/2020)), Certified 4.0 (07/2020-07/2021), and a fifth one is currently underway (Certified 5.0 (07/2021-07/2022) (Diversity Lab 2017, 2018, 2019, 2020, 2021). Each generation relies on the application of the 30% rule. None considers the impact of supply-demand dynamics on the potential pools of candidates. Shortages prevent the formation of candidate pools consisting of 30% women or underrepresented minorities and make it difficult or impossible to operate the Mansfield Rule.

### C. Is the Mansfield moving the needle

When it comes to measuring the extent to which interventions designed to reduce bias and increase diversity work or do not work, there is very little research that sheds light on these efforts and their outcomes.[25] The approach taken engages natural experiments, which were useful for assessing interventions intended to remedy discrimination and inequality in employment, housing, and other economic or social resources.[26]

In the spotlight as a result of a recent Nobel Prize, natural experiments offer an effective empirical approach to examining groups and behaviors.[27] This method is notable for its capacity to address important questions that cannot be approached through Random Control Trial.

---

[24] DIVERSITY LAB, *44 Law Firms Pilot Version of Rooney Rule to Boost Diversity in Leadership Ranks*, (2017).
[25] Frank Dobbin & Alexandra Kalev, *Why Doesn't Diversity Training Work? The Challenge for Industry and Academia*, 10 ANTHROPOLOGY NOW 48 (2018); Frank Dobbin, Alexandra Kalev & Erin Kelly, *Diversity Management in Corporate America*, 6 CONTEXTS 21 (2007); Alexandra Kalev, Frank Dobbin & Erin Kelly, *Best Practices or Best Guesses? Assessing the Efficacy of Corporate Affirmative Action and Diversity Policies*, 71 AM SOCIOL REV 589 (2006).
[26] Fuchs-Schündeln and Hassan, *supra* note 10.
[27] David Card & Alan B. Krueger, *Minimum Wages and Employment: A Case Study of the Fast-Food Industry in New Jersey and Pennsylvania*, 84 THE AMERICAN ECONOMIC REVIEW 772 (1994).

Empirical investigation grounded in natural experiment methodology has shown that the Mansfield Rule is not moving the needle for women and minorities as reported.[28] The data shows a natural growth trend that was taking place across the legal industry and similarly impacted firms that applied Mansfield and those that did not.

Firms committed to DEI prior to Mansfield already had diverse populations and continued to make progress with their numbers related to women and minorities in leadership positions.[29] They experienced growth rates consistent with industry trends, whether they applied the Mansfield Rule or not. This includes growth with respect to underrepresented groups.[30] The Mansfield Rule did not significantly affect the rate at which diversity increased.

Applying the Mansfield Rule did not observably increase—through hiring, promotion, election, or appointment—the population of women and minorities in the promised roles[31]. These roles include equity partner, lateral partner, practice group or office leader; membership executive, management, nomination, promotion, and compensation committees, and board of directors. Also included are chairperson and governance roles and participation in formal client pitches.

Detrimentally, low numbers of supply can force hiring choices that appear driven by demographics. Candidates hired under the Mansfield Rule can face a backlash of resentment or doubt regarding their qualifications, as described by economists Fershtman and Pavan in their recent article.

Finally, it is also important to note that Diversity Lab's claims that the Mansfield Rule influences law firms' diversity are based on their interpretation of the dataset from the MCCA.[32] These data are collected by MCCA on law firms' workforces. However, the same MCCA dataset has also been used in my article in which I conclude the Rule has no effect.[33] Furthermore, even the owner of the MCCA data set concluded after examining their own data that "the Mansfield rule certification does not have a direct or noticeable impact on improving diversity."[34]

### III. Methodology

This article aims to establish whether the supply of lawyers exists to meet the demand of leadership positions and provide 30% women and minorities in corresponding candidate pools. It seeks to establish if market equilibrium exists at the job level for counsel and partners (non-equity and equity). This model categorizes the population in general, then looks at two population sub-groups of lawyers: white Caucasian female and minorities (female and male).

It establishes the size of the population of available lawyers and defines the demand corresponding to the job level. It then determines if equilibrium exists, or not, in the supply and demand. In a nutshell, it reveals if there could be enough candidates for a job level, and in case of shortage or surplus, it identifies the percentage of the population above or below the equilibrium.

There are three possible outcomes in a supply-demand analysis:

---

[28] Cecchi Dimeglio, *supra* note 8.
[29] *Id.*
[30] *Id.*
[31] *Id.*
[32] See Mansfield Rule "Early Adopters" Show Significant Diversity Growth—and Outpace Legal Industry—In Critical Leadership Roles, DIVERSITY LAB (Apr. 15, 2021), https://www.diversitylab.com/pilot-projects/mansfield-rule-early-adopter-firm-results/.
[33] See Cecchi-Dimeglio, supra note 8 at 247.
[34] Elizabeth Olson, Law Firms Struggle With Their Own Rooney Rule on Diversity, BUS. INSIDER (Dec. 8, 2022, 11:01 AM), https://www.businessinsider.com/law-firms-hiring-racial-diversity-blacks-women-rooney-nfl-2022-12.

- an equilibrium between the demand and supply (at 0),
- a shortage between the demand and supply (below 0),
- or a surplus between the demand and supply (above 0).

The demand side of a specific job level is based on total lawyers available (analysis for white Caucasian female /minorities female and male), based on lawyers that can fulfil this role either by being recruited laterally or by being promoted into this position.

The availability side is based on total lawyers available (analysis per gender (female and male) and per race (white Caucasian/minorities), based on lawyers leaving their positions, which includes the level of attrition and promotion for this position.

The supply and demand model has been established as follows. The lawyer's level $X$ at a time $t$ is defined as $X_t$ and contains the following levels: associates (junior associate, mid associate, senior associate) $\{A(j)_t, A(m)_t, A(s)_t\}$, counsels $\{C_t\}$, non-equity partners $\{Neq_t\}$ and equity partners $\{Eq_t\}$.

The main quantities used to derive the supply per levels are:
- $\mathcal{L}at(X_t)$: Lateral hiring employee per group
- $\mathcal{R}(X_t)$: Retirement per category
- $\mathcal{R}et(X_t)$: Retention per category
- $\mathcal{R}eo(X_t)$: Number of senior associates which was not included in the partner promotion class
- $\mathcal{A}tt(X_t)$: Define the number of attrition of lawyer per group
- $\mathcal{P}(X_t|X_t')$: Lawyer promoted per group, where $X$ is the new position and $X'$ is the old position reached by the lawyer.

Therefore, the relationship between population is for each level $X_t$:

**Associates** $(A(j)_t, A(m)_t, A(s)_t)$ for an ordered associate level $x \in \{j, m, s\}$, with a promotion rate $p_i$:

$$A(x)_{t+1} = A(x)_t - \mathcal{A}tt(A(x))_t + \sum_{i \in \{j,m,s\}|i<x} p_i \cdot \mathcal{R}et(A(x))_t - \mathbb{I}_{(x=s)} * \mathcal{P}(C_t, Neq_t, Ep_t|A(s)_t)$$

**Counsel** $(C_t)$
$$C_{t+1} = C_t + \mathcal{L}at(C_t) + \mathcal{P}(C_t|A(s)_t) - \mathcal{P}(Neq_t, Ep_t|C_t) - \mathcal{A}tt(C_t) - \mathcal{R}(C_t)$$

**Non-equity partner** $(Neq_t)$
$$Neq_{t+1} = Neq_t + \mathcal{L}at(Neq_t) + \mathcal{P}(Neq_t|C_t, A(s)) - \mathcal{P}(Ep_t|Neq_t) - \mathcal{A}tt(Neq_t) - \mathcal{R}(Neq_t)$$

**Equity partner** $(Eq_t)$
$$Ep_{t+1} = Ep_t + \mathcal{L}at(Ep_t) + \mathcal{P}(Ep_t|C_t, A(s), Neq_t) - \mathcal{A}tt(Ep_t) - \mathcal{R}(Ep_t)$$

The following steps are taken in order to establish the supply and demand of lawyers in leadership position which include counsel and partner (non-equity and equity partner). First, the model establishes the demand which is composed of the share of positions to be filled in this job level. Second, the model determines the population that needs to be available, which considers all the lawyers available in the market to fulfil this job level.

For a population $x$, let's define $Y_+$ the demand:
$$Y_+(x) = \mathcal{P}(x|x_{old}) + \mathcal{L}at(x)$$
And the available $Y_-$
$$Y_-(x) = \mathcal{A}tt(x) + \mathcal{P}(x_{new}|x) - \mathcal{R}eo(x) - \mathcal{R}(x)$$

The demand and availability per job level can be calculated for the counsel and the partner populations for a specific subgroup $m$ of the population. The available group is defined as $Y_-^m(x)$ where $m$ is the population subgroup. Respectively, the model allows definition of the overall demand as $Y_+^\Omega(x)$ as the overall demand per race (white Caucasian/minorities) subgroup.

The main quantities used to derive the supply per levels are:
$$\mathbb{R}_{30\%}(X^m) = \frac{\sum^m Y_-^m}{0.3(Y_+^\Omega)} - 1,$$

$Y_+^\Omega$ represents the demand of the population necessary to have in the supply under the 30% Mansfield rule condition.

$\sum^m Y_-^m$ is the available amount of the subgroup population (♀white or minority ♂&♀) under the 30% Mansfield rule condition.

### IV. Descriptive Data

The dataset is compiled from the AML dataset and complemented with publicly available US Census Bureau, US Bureau of Labor Statistics, and American Bar Association data. The data covers 2018, 2019, 2020 and 2021. This article presents data for the year 2021, as they are the most recent and as the result remain almost the same for the other years, without significant change.

The dataset consists of 327 law firms with a distribution per size of law firm as follows: (1) 251-500 lawyers (155 firms), (2) 501-750 lawyers (62 firms), and (3) more than 750 lawyers (110 firms) (*Table 1*). Overall, the distribution per size of law firm in the dataset is as follows: (1) 251-500 lawyers (47%), (2) 501-750 lawyers (19%), and (3) more than 751 lawyers (34%) (*Table 1*).

**Table 1: Job level distribution**

| Size | All (N = 327) | 251-500 (N = 155 (47%)) | 501-750 (N = 62 (19%)) | More than 751 (N = 110 (34%)) |
|---|---|---|---|---|
| N | | % of the total | % of the total | % of the total | % of the total |
| Total (Distribution) | 248628 (100%) | 100% | 23% | 15% | 62% |
| Associate | 112089 | 45% | 37% | 47% | 48% |
| Partner | 103476 | 42% | 49% | 39% | 40% |
| Counsel | 33063 | 13% | 14% | 14% | 13% |
| Non-equity partner (NE) | 27364 | 11% | 18% | 10% | 9% |
| Equity partner (E) | 76112 | 31% | 31% | 29% | 31% |
| Leadership | 136539 | 55% | 63% | 53% | 52% |

The dataset is divided per gender (women and men), and per race (white Caucasian and all minority groups (African American/Black, Alaska Native/American Indian, Asian, Hispanic/Latinx, Native Hawaiian/Pacific Islander, Multiracial), (LGBTQ±, and attorneys with disabilities as recognized by American law under the Americans with Disabilities Act (ADA))).

Table 2 shows the demographic repartition of the population total by size of firm, per job level, leadership type (counsel, Partner (Non-Equity and Equity)), and by the distribution in percentage of female and male as well as white Caucasian and minority populations.

The total population of the data set is 248,628 lawyers. The gender distribution of this population is 90,891 female lawyers ( 37%) and 157,737 male lawyers (63%). The total white Caucasian population of the data set is 204,802 lawyers (82% of the population). The total minority population of the data set is 43,826 lawyers (18% of the population).

The total population of the data set in leadership positions (which includes counsel and partner (non-equity and equity partner)) is 136,539 lawyers (55% of the total dataset). The total gender distribution in leadership positions of the population of the data set is 38,342 female lawyers (28% of the population) and 98,197 male lawyers (72% of the population). The total white Caucasian population of the data set in leadership positions is 121,558 lawyers (89% of the leadership position population). The total minority population of the data set in leadership positions is 14,981 lawyers (11% of the leadership position population).

The total population of the data set in counsel positions is 33,063 lawyers (13% of the total dataset). The total gender distribution of the data set in counsel positions is 13,332 female lawyers (40% of the population) and 19,731 male lawyers (60% of the population). The total minority counsel population of the data set is 4,555 lawyers (14% of the counsel population). The total white Caucasian population of the data set is 28,508 lawyers (86% of the counsel population).

The total population of the data set for partner (non-equity and equity) is 103,476 lawyers (42% of the total dataset). The total gender distribution for the partner population of the data set is 25,010 female lawyers (24% of the population) and 78,466 male lawyers (76% of the population). The total minority partner population of the data set is 10,426 lawyers (10% of the population). The total white Caucasian partner population of the data set is 93,050 lawyers (90% of the partner population).

The total population of the data set for non-equity partner is 27,364 lawyers (11% of the total dataset). The total gender distribution for non-equity partner population of the data set is 8,305 female lawyers (30% of the population) and 19,059 male lawyers (70% of the population). The total minority population for non-equity partner of the data set is 3,188 lawyers (12% of the population). The total white Caucasian population of the data set is 24,176 lawyers (88% of the non-equity partner population).

The total population of the data set for equity partner is 76,112 lawyers (31% of the total dataset). The total gender distribution for the equity partner population of the data set is 16,705 female lawyers (22% of the population) and 59,407 male lawyers (78% of the population). The total minority population of equity partner in the data set is 7,238 lawyers (10% of the population). The total white Caucasian population of equity partner in the data set is 68,874 lawyers (90% of the population).

Table 2: Demographic repartition

| Population | N | % of the total | Female % pop. | Male % pop. | Minorities % pop. | White % pop. |
|---|---|---|---|---|---|---|
| *Law firms: All (N = 327)* | | | | | | |
| Total | **248628** | **100%** | **37%** | **63%** | **18%** | 82% |
| Associate | 112089 | 45% | 47% | 53% | 26% | 74% |
| Counsel | 33063 | 13% | 40% | 60% | 14% | 86% |
| Partner | 103476 | 42% | 24% | 76% | 10% | 90% |
| Non-equity partner | 27364 | 11% | 30% | 70% | 12% | 88% |
| Equity partner | 76112 | 31% | 22% | 78% | 10% | 90% |
| **Leadership** | **136539** | **55%** | **28%** | **72%** | **11%** | **89%** |
| *Law firm with more than 751 lawyers (N = 110 (34%))* | | | | | | |
| Total | **154687** | | **37%** | **63%** | **19%** | **81%** |
| Associate | 73760 | 48% | 47% | 53% | 27% | 73% |
| Counsel | 19635 | 13% | 40% | 60% | 14% | 86% |
| Partner | 61292 | 40% | 24% | 76% | 11% | 89% |
| Non-equity partner | 13660 | 9% | 31% | 69% | 12% | 88% |
| Equity partner | 47632 | 31% | 22% | 78% | 10% | 90% |
| **Leadership** | **80927** | **52%** | **28%** | **72%** | **12%** | **88%** |
| *Law firm with between 501-750 lawyers (N = 62 (19%))* | | | | | | |
| Total | **37721** | | **37%** | **63%** | **18%** | **82%** |
| Associate | 17549 | 47% | 47% | 53% | 26% | 74% |
| Counsel | 5455 | 14% | 43% | 57% | 14% | 86% |
| Partner | 14717 | 39% | 24% | 76% | 9% | 91% |
| Non-equity partner | 3865 | 10% | 30% | 70% | 11% | 89% |
| Equity partner | 10852 | 29% | 22% | 78% | 9% | 91% |
| **Leadership** | **20172** | **53%** | **29%** | **71%** | **11%** | **89%** |
| *Law firm with between 251-500 lawyers (N = 155 (47%))* | | | | | | |
| Total | **56220** | | **34%** | **66%** | **14%** | **86%** |
| Associate | 20780 | 37% | 45% | 55% | 21% | 79% |
| Counsel | 7973 | 14% | 38% | 62% | 12% | 88% |
| Partner | 27467 | 49% | 24% | 76% | 9% | 91% |
| Non-equity partner | 9839 | 18% | 30% | 70% | 11% | 89% |
| Equity partner | 17628 | 31% | 21% | 79% | 8% | 92% |
| **Leadership** | **35440** | **63%** | **27%** | **73%** | **9%** | **91%** |

## V. Results

### A. Law firms larger than 751 lawyers

#### 1. State of supply and demand for the leadership (counsel and partner (non-equity and equity)) population

Table 2a shows that for firm larger than 751 lawyers, the total population in leadership (counsel and partner (non-equity and equity)) is 80,927 lawyers, with a distribution of the population at 24% counsel, and 76 % partners (divided between 17% non-equity partners and 59% equity partners).

The total demand for the population in leadership position in firms larger than 751 lawyers is 12,358 individuals (15% of the total lawyer leadership population). The demand within the leadership position is as follows: 21% of counsel and 79% for partner (divided between 22% non-equity partners and 57% equity partners).

The 30% proportional demand for the population in leadership position in firms larger than 751 lawyers is 3,708 individuals. The demand within the leadership positions is as follows: 21% of counsel and 79% for partner (divided between 22% non-equity partners and 57% equity partners). The 30% demand represents 13% of counsel and 26% of partners, composed of 20% of the non-equity partners and 15% of the equity partners.

The total population of leadership positions in firms larger than 751 lawyers available is 9,397 individuals (12% of the total lawyer leadership population). The available population within the leadership position is as follows: 29% of counsel and 71% for partner (divided between 37% non-equity partners and 34% equity partners). The available population represents 14% of the counsel and 14% of the partners composed of 26% of the non-equity partners and 5% of the equity partners.

**Table 2a: Supply chain for law firms larger than 751 lawyers**

| | Counsel | Partner (NE+E) | Non-equity partner (NE) | Equity partner (E) | Total |
|---|---|---|---|---|---|
| Leadership population (Counsels & Partners (incl. NE& E)) | 19635 | 61292 | 13660 | 47632 | 80927 |
| Population distribution | 24% | 76% | 17% | 59% | 100% |
| Demand | 2623 | 9735 | 2741 | 6994 | 12358 |
| Demand pop. distribution | 21% | 79% | 22% | 57% | 100% |
| Demand in proportion of the leadership population | 13% | 16% | 20% | 15% | 15% |
| 30% of the demand | 787 | 2921 | 822 | 2098 | 3708 |
| Available | 2681 | 6716 | 3494 | 3222 | 9397 |
| Available pop. distribution | 29% | 71% | 37% | 34% | 100% |
| Available in proportion of the leadership population | 14% | 14% | 26% | 5% | 12% |

## 2. State of the supply and demand for the white Caucasian female population

Within the proportional demand of the 30% of the population in leadership position in firms larger than 751 lawyers, the white Caucasian female population available is 2,236 individuals (24% of the total available) (Table 2b). The population available in the white Caucasian female sub-population is 783 counsels (29% of the counsel available) and 1,453 Partners (22% of the partners available) composed of 816 non-equity partners (23% of the non-equity partners available) and 637 equity partners (20% of the equity partners available))

The supply available—e.g. the capacity of the white Caucasian female population in leadership to fill the demand—is 18%. Per job level, the supply is 30% of counsel, 15% of Partners (30% non-equity partners and 9% equity partners).

There is a total shortage of 12% (=18%-30%) in the availability of white Caucasian female in leadership to fill the 30%. Nonetheless, the distribution among counsels and partner is not equal. The white Caucasian female counsel population is missing 1% of their population and the partner population is short 50% in the number of partners needed for equilibrium. However, the distribution of the shortage of white Caucasian female partners is not the same for equity and non-equity. White Caucasian female non-equity partner is only short of 1% of their population to be at equilibrium, whereas White Caucasian female equity partner is shy 70% in the number of partners to be at equilibrium for a 30% rule.

Overall, there is a need to increase the minority population by 40% to be able to realize a 30% rule threshold.

**Table 2b: White Caucasian female supply chain for law firms larger than 751 lawyers**

| | Counsel | Partner | Non-equity partner (NE) | Equity partner (E) | Total |
|---|---|---|---|---|---|
| White Caucasian female available | 783 | 1453 | 816 | 637 | 2236 |
| Proportion of Available | 29% | 22% | 23% | 20% | 24% |
| Fill capacity | 30% | 15% | 30% | 9% | 18% |
| Short % to reach 30% of the demand | 0% | 15% | 0% | 21% | 12% |
| Short % of white Caucasian female to reach 30% demand | -1% | -50% | -1% | -70% | -40% |

## 3. State of the supply and demand for the minority population

Within the proportional demand of 30% of the population in leadership position in firms larger than 751 lawyers, the minority population available is 1,434 individuals and represents 15% of the total population available (Table c). The population available in the minority sub-population is 431 counsels (16% of the counsel available) and 1,003 Partners (15% of the partners available), composed of 522 non-equity partners (15% of the non-equity partners available) and 481 equity partners (15% of the equity partners available)).

The supply available – e.g. the capacity- of the minority population in leadership to fill the demand is 15%, composed of 16% of counsel, 15% of Partners (15% non-equity partners and 15% equity partners).

There is a total shortage of 18 % (=12%-30%) in the availability of the minority population in leadership to fill the 30% requirement. However, the distribution among counsels and partner is not equal. The minority counsel population is short of 45% of their population and the Partner population is shy of 66% in the number of partners to be at equilibrium. Nevertheless, the distribution of the shortage of minority partners is not equally spread between non-equity and equity partners. Minority non-equity partner is shy of 37% of their population whereas Minorities equity partner is short of 77% in the number of partners to be at equilibrium meeting a 30% rule.

Overall, there is a need to increase the minority population by 61% to be able to realize a 30% rule threshold.

**Table 2c: Minority supply chain for law firms larger than 751 lawyers**

|   | Counsel | Partner | Non-equity partner (NE) | Equity partner (E) | Total |
|---|---|---|---|---|---|
| Minorities (female & male) available | 431 | 1003 | 522 | 481 | 1434 |
| Proportion of Available | 16% | 15% | 15% | 15% | 15% |
| Fill capacity | 16% | 10% | 19% | 7% | 12% |
| Short % to reach 30% of the demand | 14% | 20% | 11% | 23% | 18% |
| Short % for minorities (female & male) to reach 30% demand | -45% | -66% | -37% | -77% | -61% |

Graphic 1 represents visually the statistical numbers presented above in tables 2a, 2b, 2c, and visually represents the supply and demand chain for law firms with more than 751 lawyers

**Graphic 1: Supply chain for law firms larger than 751 lawyers**

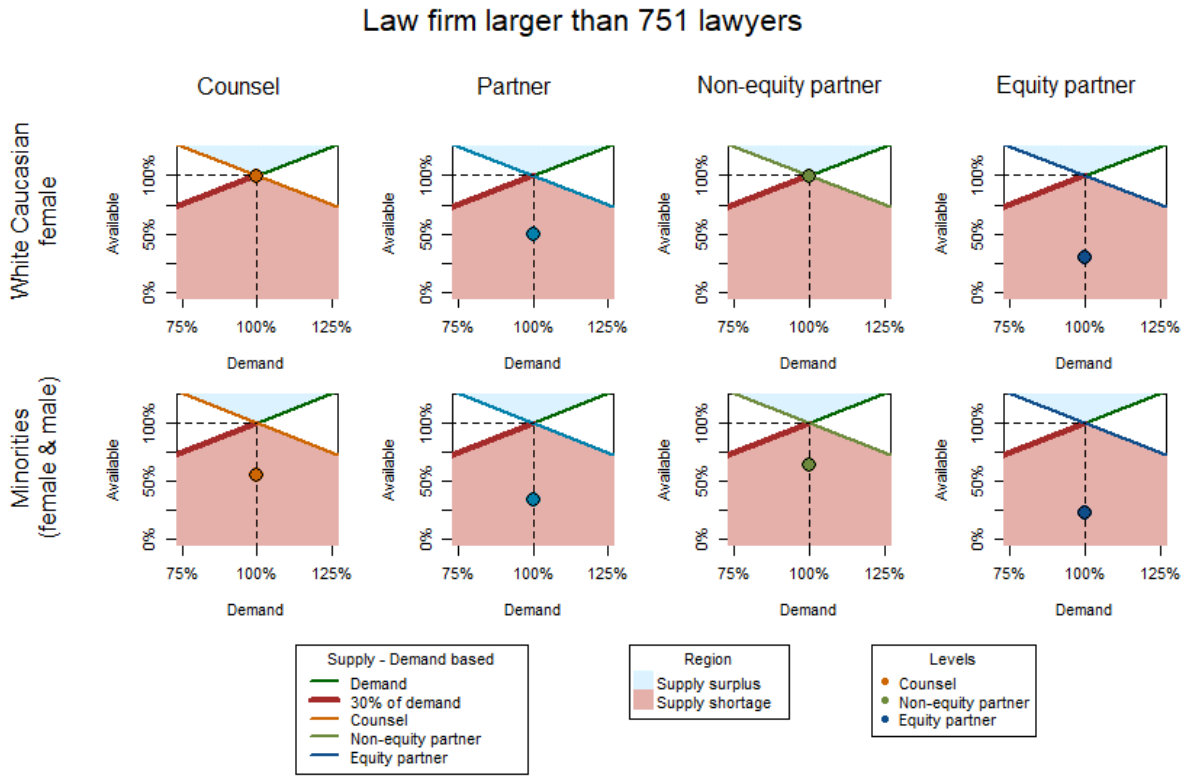

B. **Law firms between 501-750 lawyers**

1. **State of the supply and demand for the leadership (counsel and partner (non-equity and equity)) population**

Table 3a shows that for firms between 501-750 lawyers, the total population in leadership (counsel and partner (non-equity and equity)) is 20,172 lawyers, with a distribution of the population at 27% of counsel, and 73 % of partners (divided between 19 % non-equity partners and 54% equity partners).

The total demand for the population in leadership positions in firms between 500-751 lawyers is 3,456 individuals (17% of the total lawyer leadership population). The demand within the leadership position is as follows: 33% of counsel and 67% for partner (divided between 18 % non-equity partners and 49% equity partners).

The 30% proportional demand for the population in leadership positions in firms between 500-751 lawyers is 1,037 individuals. The demand within the leadership position is as follows: 33% of counsel and 67% for partner (divided between 18% non-equity partners and 49% equity partners). The 30% demand represents 21% of the counsel and 16% of the partners, composed of 16% of the non-equity partners and 16% of the equity partners.

The total population in leadership positions in firms between 500-751 lawyers is 3,258 individuals (16% of the total lawyer population). The available population within the leadership positions is as follows: 28% of counsel and 72% for partner (divided between 19 % non-equity partners and 53% equity partners). The available represents 17% of the counsel and 22% of the partners composed of 16% of the non-equity partners and 12% of the equity partners.

**Table 3a: Supply chain for law firms with 501 to 750 lawyers**

|  | Counsel | Partner | Non-equity partner (NE) | Equity partner (E) | Total |
|---|---|---|---|---|---|
| Leadership population (Counsels & Partners (incl. NE& E)) | 5455 | 14717 | 3865 | 10852 | 20172 |
| Population distribution | 27% | 73% | 19% | 54% | 100% |
| Demand | 1144 | 2312 | 618 | 1694 | 3456 |
| Demand pop. distribution | 33% | 67% | 18% | 49% | 100% |
| Demand in proportion of the leadership population | 21% | 16% | 16% | 16% | 17% |
| 30% of the demand | 343 | 694 | 185 | 508 | 1037 |
| Available | 913 | 2345 | 625 | 1720 | 3258 |
| Available pop. distribution | 28% | 72% | 19% | 53% | 100% |
| Available in proportion of the leadership population | 17% | 22% | 16% | 12% | 16% |

## 2. State of the supply and demand for the white Caucasian female population

Within the proportional demand of the 30% of the population in leadership positions in firms between 500-751 lawyers, the white Caucasian female population available is 751 individuals and represents 23% of the total available (Table 3b). The population available in the white Caucasian female sub-population is 270 counsels (30% of the counsel available) and 481 Partners (21% of the partners available), composed of 162 non-equity partners (26% of the non-equity partners available) and 319 equity partners (19% of the equity partners available).

The supply available—e.g. the capacity of the white female population in leadership to fill the demand—is 22%. Per position it is 24% of counsel, 21% of Partners (26% non-equity partners and 19% equity partners).

There is a total shortage of 8% (=22%-30%) in the availability of white Caucasian females in leadership to fill the 30%. However, the distribution among counsels and partner is not equal.

The white Caucasian female counsel population is short 21% and the partner population is short of 31% of the numbers needed to be at equilibrium. However, the distribution of the shortage amount of white Caucasian female partner is not the same for equity and non-equity. White Caucasian female non-equity partner is short of 12% of the population needed to be at equilibrium whereas White Caucasian female equity partner is shy of 37% in the number of partners to be at equilibrium meeting a 30% rule.

Overall, there is a need to increase the minority population by 28% to be able to realize a 30% rule threshold.

**Table 3b: White Caucasian female supply chain for law firms with 501 to 750 lawyers**

|  | Counsel | Partner | Non-equity partner (NE) | Equity partner (E) | Total |
|---|---|---|---|---|---|
| White Caucasian female available | 270 | 481 | 162 | 319 | 751 |
| Proportion of Available | 30% | 21% | 26% | 19% | 23% |
| Fill capacity | 24% | 21% | 26% | 19% | 22% |
| Short % to reach 30% of the demand | 6% | 9% | 4% | 11% | 8% |
| Short % of white Caucasian female to reach 30% demand | -21% | -31% | -12% | -37% | -28% |

### 3. State of the supply and demand for the minority population

Within the proportional demand of the 30% of the population in leadership position in firms between 500-751 lawyers, the minority population available is 443 individuals (14% of the total available) (Table 3c). The population available in the minority sub-population is 169 counsels (19% of the counsel available) and 273 Partners (12% of the partners available) composed of 80 non-equity partners (13% of the non-equity partners available) and 193 equity partners (11% of the equity partners available).

The supply available – e.g. the capacity- of the minority population in leadership to fill the demand is 13%, composed of 15% of counsel, 12% of Partners (13% non-equity partners and 11% equity partners).

There is a total shortage of 17% (=13%-30%) in the availability of the minority population in leadership to fill the 30%. However, the distribution among counsels and partner is not equal. The minority population of counsel is shy of 51% of their population to be and the Partner population is short 61% in the number of partners to be at equilibrium. Nevertheless, the distribution of the shortage amount is not the same for equity and non-equity minorities partners. Minorities non-equity partner is short 57% of their population, whereas Minorities equity partner is shy of 62% in the number of partners to be at equilibrium meeting a 30% rule.

Overall, there is a need to increase the minority population by 57% in order to realize a 30% rule threshold.

Table 3c: Minorities supply chain for law firms with 501 to 750 lawyers

|  | Counsel | Partner | Non-equity partner (NE) | Equity partner (E) | Total |
|---|---|---|---|---|---|
| Minorities (female & male) available | 169 | 273 | 80 | 193 | 443 |
| Proportion of Available | 19% | 12% | 13% | 11% | 14% |
| Fill capacity | 15% | 12% | 13% | 11% | 13% |
| Short % to reach 30% of the demand | 15% | 18% | 17% | 19% | 17% |
| Short % for minorities (female & male) to reach 30% demand | -51% | -61% | -57% | -62% | -57% |

Graphic 2 represents visually the statistical numbers presented above in tables 3a, 3b, 3c, and visually represent the supply and demand chain for law firms with 501 to 750 lawyers

**Graphic 2: Supply chain for law firms with 501 to 750 lawyers**

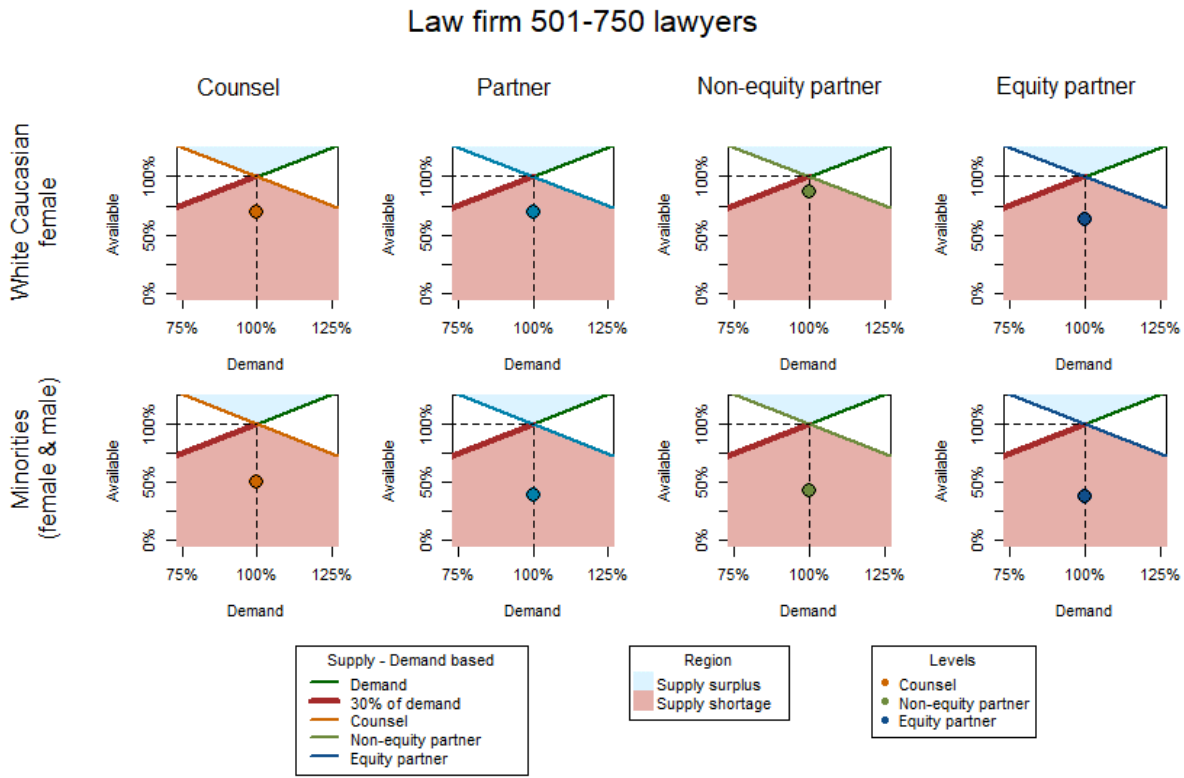

C. **Firm between 251-500 lawyers**

1. **State of the supply and demand for the leadership (counsel and partner (non-equity and equity)) population**

Table 4a shows that for firm from 251 to 500 lawyers, the total population in leadership (counsel and partner (non-equity and equity)) is 35,440 lawyers with a distribution of the population at 22% of counsel, and 78% of partners (divided between 28% non-equity partners and 50% equity partners).

The total demand for the population in leadership position in firms from 251 to 500 lawyer is 4,396 individuals which representing 12% of the total lawyer leadership population. The demand within the leadership position is as follows: 24% of counsel and 76% for partner (divided between 34% non-equity partners and 42% equity partners).

The 30% proportional demand for the population in leadership position in firms from 251 to 500 lawyer is 1,319 individuals. The demand within the leadership position is as follows: 24% of counsel and 76% for partner (divided between 34% non-equity partners and 42% equity partners). The demand represents 13% of the counsel and 12% of the partners composed of 15% of the non-equity partners and 10% of the equity partners.

The total population in leadership position in firms from 251 to 500 lawyer available is 3,954 individuals which representing 11% of the total lawyer population. The available population within the leadership position is as follows: 31% of counsel and 69% for partner (divided between 44 % non-equity partners and 25% equity partners). The available represents 15% of the counsel and 16% of the partners composed of 18% of the non-equity partners and 4% of the equity partners.

**Table 4a: Supply chain for law firms with 251 to 500 lawyers**

|  | Counsel | Partner | Non-equity partner (NE) | Equity partner (E) | Total |
|---|---|---|---|---|---|
| Leadership population (Counsels & Partners) | 7973 | 27467 | 9839 | 17628 | 35440 |
| Population distribution | 22% | 78% | 28% | 50% | 100% |
| Demand | 1072 | 3324 | 1477 | 1848 | 4396 |
| Demand pop. distribution | 24% | 76% | 34% | 42% | 100% |
| Demand in proportion of the leadership population | 13% | 12% | 15% | 10% | 12% |
| 30% of the demand | 322 | 997 | 443 | 554 | 1319 |
| Available | 1209 | 2745 | 1742 | 1003 | 3954 |
| Available pop. distribution | 31% | 69% | 44% | 25% | 100% |
| Available in proportion of the leadership population | 15% | 16% | 18% | 4% | 11% |

### 2. State of the supply and demand for the white Caucasian female population

Within the proportional demand of the 30% of the population in leadership position in firms from 251 to 500 lawyer the white female population available is 966 individuals and represents 24% of the total available (Table 4b). The population available in the white female sub-population is 318 counsels (26% of the counsel available) and 648 Partners (24% of the partners available) composed of 433 non-equity partners (25% of the non-equity partners available) and 215 equity partners (21% of the equity partners available)).

The supply available – e.g. the capacity-of the white female population in leadership to fill the demand is 24%. Per position it is 26% of counsel, 24% of Partners (25% non-equity partners and 21% equity partners).

There is a total shortage of 8%( =22%-30%) in the availability of white female in leadership to fill the 30%. However, the distribution among counsels and partner is not equal. The white female Caucasian counsel population is at equilibrium. However, the Partner population is short of 11% in the number of partners to be at equilibrium. Nevertheless, the distribution of the shortage of white Caucasian female is not the same for equity and non-equity partners. White Caucasian female non-equity partner is only short of 1% of the population needed to be at equilibrium whereas White Caucasian female equity partner is shy of 18% in the number of partners to arrive at equilibrium meeting a 30% rule.

Overall, there is a need to increase by 73% in the minority population to be able to realize a 30% rule threshold.

**Table 4b: White Caucasian female supply chain for law firms with 251 to 500 lawyers**

|  | Counsel | Partner | Non-equity partner (NE) | Equity partner (E) | Total |
|---|---|---|---|---|---|
| White Caucasian female available | 318 | 648 | 433 | 215 | 966 |
| Proportion of Available | 26% | 24% | 25% | 21% | 24% |
| Fill capacity | 30% | 19% | 29% | 12% | 22% |
| Short % to reach 30% of the demand | 0% | 11% | 1% | 18% | 8% |
| Short % of white Caucasian female to reach 30% demand | -1% | -35% | -2% | -61% | -27% |

### 3. State of the supply and demand for the minority's population

Within the proportional demand of the 30% of the population in leadership position in firms from 251 to 500 lawyers, the minority population available is 483 individuals and represents 12% of the total available (Table 4c). The population available in the minority sub-population is 196 counsels (16% of the counsel available) and 287 Partners (10% of the partners available) composed of 184 non-equity partners (11% of the non-equity partners available) and 103 equity partners (10% of the equity partners available)).

The supply available – e.g. the capacity- of the minority population in leadership to fill the demand is 11%, composed of 18% of counsel, 9% of Partners (12% non-equity partners and 6% equity partners).

There is a total shortage of 19% (=11%-30%) in the availability of the minority population in leadership to fill the 30%.

However, the distribution among counsels and partner is not equal. The minority population counsel is short of 39% of their population to be and the Partner population is missing 71% in the number of partners. However, the distribution of the shortage of minorities partner is not equally distributed between equity and non-equity partners. Minorities non-equity partner is shy of 59% of the population needed to be at equilibrium whereas Minorities equity partner is short of 81% in the number of partners to be at equilibrium meeting a 30% rule.

Overall, there is a need to increase by 63% in the minority population to be able to realize a 30% rule threshold.

Table 4c: Minorities supply chain for law firms with 251 to 500 lawyers

|  | Counsel | Partner | Non-equity partner (NE) | Equity partner (E) | Total |
|---|---|---|---|---|---|
| Minorities (female & male) available | 196 | 287 | 184 | 103 | 483 |
| Proportion of Available | 16% | 10% | 11% | 10% | 12% |
| Fill capacity | 18% | 9% | 12% | 6% | 11% |
| Short % to reach 30% of the demand | 12% | 21% | 18% | 24% | 19% |
| Short % for minorities (female & male) to reach 30% demand | -39% | -71% | -59% | -81% | -63% |

Graphic 3 represents visually the statistical numbers presented above in tables 4a, 4b, 4c, and visually represent the supply and demand chain for law firms with 250 to 500 lawyers

Graphic 3: Supply chain for law firms with 251 to 500 lawyers

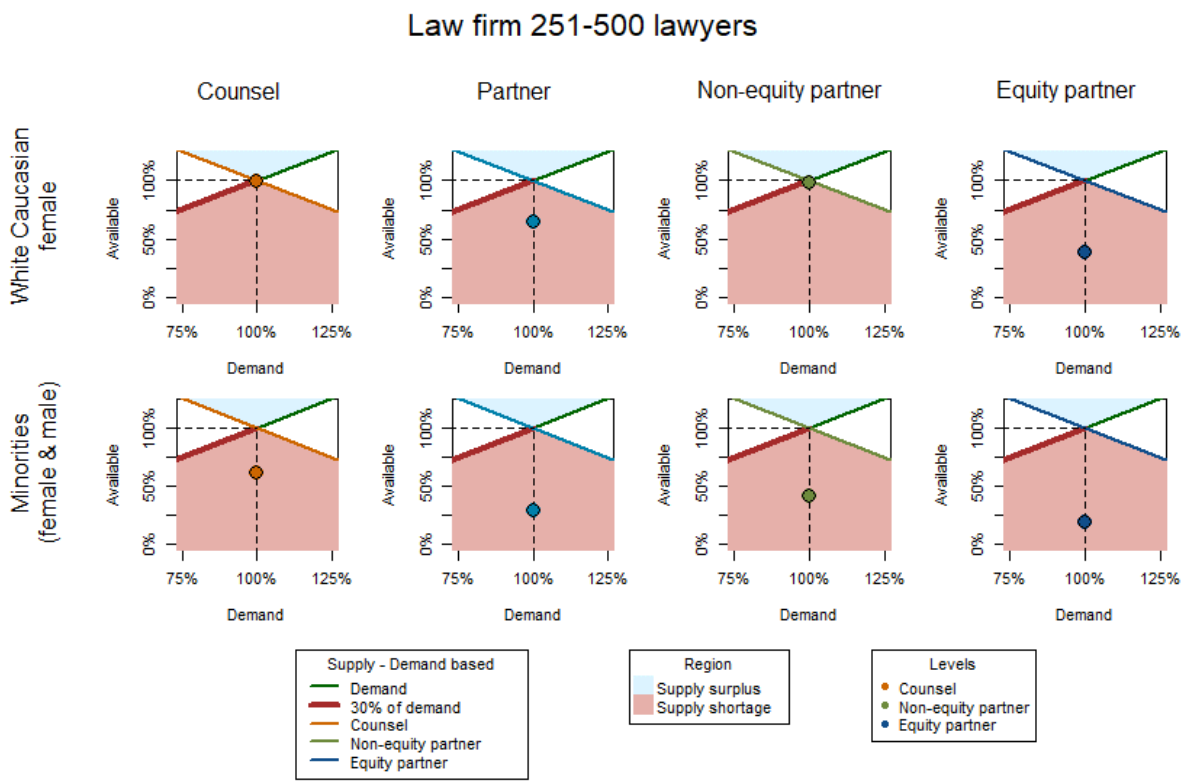

## VI. Discussion

Although recruiting lawyers (white Caucasian female or minorities (female and male)) into leadership positions is dependent upon supply and availability within that population, this article is one of the first to present a supply-demand analysis of the legal leadership population (counsel and partner (non-equity and equity)).

Per the findings, demand at law firms larger than 751 lawyers is 15%, 17% at firms between 501-750, and 12% at firm between 251-500. However, supply for law firms larger than 751 lawyers is 12%, 16% for firms between 501-750, and 11% for firms between 251-500.

These findings indicate a shortage. Equilibrium is not reachable, despite an 1% annual grow of the legal population.

Overall, the data shows there is currently a ceiling in the supply of 18% for law firms larger than 751 lawyers, 22% for firms between 501-750, and 22% for firms between 251-500.

Furthermore, a different supply ceiling exists for counsel and partners (non-equity and equity): 15% for partners and 30% for counsel at law firms larger than 751 lawyers, 21% for partners and 24% for counsel at firms between 501-750, and 19% for partners and 30% for counsel at firms between 251-500.

The findings conclude that there is insufficient supply to fuel the Mansfield rule's 30% engine. Mathematically, the Mansfield rule is not even achievable, given the shortage of supply in the population available and demand in leadership (counsels and partners (non-equity and equity)).

Mansfield posits that having women, minorities, LGBTQ±, and individuals with disabilities as 30% of the pipeline is sufficient to increase their numbers among those who are hired or included. Diversity Lab claims that the "Mansfield rule is based on decades of science and data" and that "transparency and accountability are baked into the structured certification process, with the requirements evolving and getting tougher every year."[35] These mathematically sound findings demonstrate that there is no 30% pipeline, the core mechanism of the Mansfield dynamic.

Though the Diversity Lab argument may be scientifically based, it is not applicable in the existing supply-demand landscape. If one applies the logic of the study referred above by the Mansfield rule and create a pool of 4 candidates in which 30% should be diverse and as such are more likely to be hired. The 30% rule of a pool of 4 candidates is equal to 1.2.

This reasoning implies that in a scenario in which 5 pools of 4 candidates exist, there is only 4 pools of candidate with 1 diverse candidate (women or minorities), and as such the situation is similar that in the paper cited in which this candidate has 0% likelihood of being hired; and 1 pool of candidate with 2 diverse candidates (women and/or minorities) in which these candidates have 50% likelihood of being hired.

As a result, there is a 10% chance (1 pool of 5 has one possibility to be hire and when being part of the pool that is recruited the candidate has 50% chance) for the diverse candidate to be hired, which is very far from the 30% theoretically given by the Mansfield. Furthermore, the deviation in expected probability referred by the scientific article cited by the Diversity lab can't not even be realized in the Mansfield situation.

As a result, the Mansfield rule is forcing law firms into a behavior of "force choice of candidate" and reinforce tokenism of the very same population that is said to help growing in leadership.

---

[35] David R. Hekman et al., *Does Diversity-Valuing Behavior Result in Diminished Performance Ratings for Non-White and Female Leaders?*, 60 AMJ 771 (2017); Stefanie K. Johnson, David R. Hekman & Elsa T. Chan, *If There's Only One Woman in Your Candidate Pool, There's Statistically No Chance She'll Be Hired*, Harvard Business Review, 2016, https://hbr.org/2016/04/if-theres-only-one-woman-in-your-candidate-pool-theres-statistically-no-chance-shell-be-hired (last visited Sep 26, 2022).

This research leads to other avenue to be further investigate such as if the Mansfield Rule impact the behavior of individuals in their lateral move and accelerate the lateral movements of women and diverse candidate to join law firms that are certified compared to non-certified one.

The findings call for structural reform to sustain the future of increasing the number of female and minorities in leadership position. The result also debunks the myth establish by the 30% Mansfield rule as a tool enhancing diversity in the legal profession especially in leadership.

The "small-N problem" is real in the legal profession especially in leadership and the ability to further develop role model is depending on the pipeline and the leak occurring along the path before one as the ability to access leadership position. Economic disruptions occur more often than realized, and these event soften trigger individual reevaluation of career and life decisions. Rethinking and revised decisions can have long-term impact on individuals and the larger economy. Different time and career stages help shape the outcomes for individuals.

Within this scope of possibilities, actors along the pipeline process need to take action to remedy the current situation in which the legal profession finds itself. On one hand, law schools should be more inviting to diverse candidate and increase the pipeline of diverse graduates. On the other hand, law firms should take the time and responsibility to invest along the talent pipeline and create alternative work structure solution in order to avoid accentuating the gap that exist in the supply and find ways to reduce it.

**ACKNOWLEDGEMENTS**

This research received no external funding.